\documentclass[letterpaper,twocolumn,10pt]{article}

\usepackage[T1]{fontenc}
\usepackage{mathptmx}
\usepackage{mathrsfs}

\usepackage{amsmath,amssymb}
\usepackage[letterpaper,top=1in,bottom=1in,left=0.75in,right=0.75in,columnsep=0.33in]{geometry}
\usepackage{graphicx}
\usepackage{booktabs}
\usepackage{array}
\usepackage{tabularx}
\usepackage[table]{xcolor}
\usepackage{enumitem}
\usepackage{titlesec}
\usepackage[font=footnotesize,labelfont=bf,labelsep=colon]{caption}
\usepackage{listings}
\usepackage{tikz}
\usetikzlibrary{arrows.meta,positioning,calc,fit,backgrounds}
\usepackage{pgfplots}
\pgfplotsset{compat=1.17}
\usepackage{url}
\usepackage[colorlinks=true,allcolors={blue!70!black}]{hyperref}
\hypersetup{pdftitle={Governing Enterprise AI Traffic at Scale: Design and Operation of a Policy-Driven, Multi-Vendor LLM Gateway on Envoy},pdfauthor={Mithun Sasidharan, Srinivasan Manoharan, Rupesh Thakur, Abhinav TK, Ashwin Hegde, Anand Gautam}}

\titleformat{\section}{\Large\bfseries\raggedright\hyphenpenalty=10000}{\thesection}{0.8em}{}
\titleformat{\subsection}{\large\bfseries\raggedright\hyphenpenalty=10000}{\thesubsection}{0.8em}{}
\titlespacing*{\section}{0pt}{1.6ex plus .4ex}{1.0ex}
\titlespacing*{\subsection}{0pt}{1.4ex plus .3ex}{0.6ex}
\newcommand{\para}[1]{\par\smallskip\noindent\textbf{#1}\hspace{0.45em}}
\setlist{nosep,leftmargin=*}
\setlist[itemize]{label=\textbullet}

\definecolor{cblue}{RGB}{31,94,168}
\definecolor{corange}{RGB}{196,109,10}
\definecolor{cgreen}{RGB}{46,125,70}
\definecolor{cgray}{RGB}{110,110,115}
\definecolor{codebg}{RGB}{245,245,245}
\definecolor{codeframe}{RGB}{200,200,200}
\definecolor{kw}{RGB}{20,40,140}
\definecolor{str}{RGB}{170,20,20}
\definecolor{cmt}{RGB}{120,120,120}

\lstdefinestyle{code}{
  basicstyle=\ttfamily\scriptsize,
  backgroundcolor=\color{codebg},
  frame=single, rulecolor=\color{codeframe},
  framesep=3pt, xleftmargin=3pt, xrightmargin=3pt,
  columns=fullflexible, keepspaces=true,
  showstringspaces=false,
  commentstyle=\color{cmt}\itshape,
  stringstyle=\color{str},
  keywordstyle=\color{kw},
  breaklines=true, breakindent=2em,
  aboveskip=4pt, belowskip=4pt,
}
\lstdefinelanguage{rego}{
  morekeywords={package,import,contains,if,some,in,not,else,true,false},
  morecomment=[l]{\#},
  morestring=[b]",
}
\lstdefinelanguage{pseudo}{
  morecomment=[l]{\#},
}
\lstdefinelanguage{yamlish}{
  morecomment=[l]{\#},
  morestring=[b]",
}

\tikzset{
  box/.style={draw, rounded corners=2pt, align=center, inner sep=3pt, line width=0.5pt},
  gbox/.style={box, draw=cgray, fill=gray!8},
  bbox/.style={box, draw=cblue, fill=blue!7},
  wbox/.style={box, draw=cblue, fill=white},
  obox/.style={box, draw=corange, fill=orange!10},
  ebox/.style={box, draw=cgreen, fill=green!7},
  arr/.style={-{Stealth[length=4pt]}, line width=0.5pt, draw=black!75},
  darr/.style={-{Stealth[length=4pt]}, line width=0.5pt, draw=black!60, dashed},
  sub/.style={font=\tiny},
}

\begin{document}

\title{\Large\bfseries Governing Enterprise AI Traffic at Scale:\\[2pt]
\large Design and Operation of a Policy-Driven, Multi-Vendor LLM Gateway on Envoy}
\author{
Mithun Sasidharan\thanks{Corresponding author: \texttt{msasidharan@paypal.com}}
\quad Srinivasan Manoharan \quad Rupesh Thakur\\
Abhinav TK \quad Ashwin Hegde \quad Anand Gautam\\[4pt]
{\small PayPal Inc.}
}
\date{}

\twocolumn[
\begin{@twocolumnfalse}
\maketitle
\end{@twocolumnfalse}
]

\section*{Abstract}
Large organizations now consume large language models (LLMs) from several vendors at once, and each vendor brings its own API dialect, credential model, streaming protocol, pricing and release cadence. Integrated directly, this heterogeneity yields credential sprawl, format lock-in, unbounded spend and inconsistent governance. We present the design and production operation of the \textbf{Cosmos LLM Gateway}, the governed entry point for enterprise LLM traffic at PayPal. The gateway is built on Envoy and splits work by latency profile: I/O-bound control decisions (authentication, credential brokering, policy and budget evaluation) run in an out-of-process external-processing service, while latency-sensitive payload work (wire-format translation, token metering and prompt-cache shaping) runs in-proxy as a Rust WebAssembly filter, so that, by default, streamed response body chunks never wait on an out-of-process call. Models, providers and prices are largely a validated data model rather than code. Governance is declarative policy evaluated by Open Policy Agent and managed through a control plane with role-scoped authoring for platform and business-unit administrators, approval workflows, scheduled activation, limited-access model releases, cohort-based pilots, shadow-mode rollout, configurable usage notifications and versioned policy-bundle distribution. For use cases that require it, a data-loss-prevention layer that combines pattern rules with an internally hosted detection model can block, redact or reversibly pseudonymize detected sensitive content before it reaches a provider. Every spend scope is a \emph{view} over per-model usage counters, so a new budget applies to spend already recorded in its window without data migration, and every governed request is designed to yield one deduplicated usage record in the analytics warehouse. As of September 2026, the gateway serves more than 12,000 developers and hundreds of business use cases, including real-time voice support with more than 2,000 concurrent calls. It handles millions of requests and roughly 200 billion tokens per day across more than 100 models from multiple providers, at 99.999\% availability to business. In a benchmark of the core request path against a synthetic upstream, it adds a median of 4\,ms and a p99 of 11\,ms of overhead at more than 2,200 requests per second on two 4-vCPU instances, with no failed requests across the tested range. We report the architecture, the policy model, an evaluation, and the lessons we would carry to any enterprise standardizing access to multiple AI providers.

\smallskip
\noindent\textbf{Keywords:} LLM gateway, AI governance, Envoy, ext\_proc, WebAssembly, policy as code, Open Policy Agent, data loss prevention, FinOps, multi-vendor routing, observability.

\section{Introduction}
Generative AI quickly became a production dependency at PayPal. Engineers use agentic coding tools daily; product teams ship AI-assisted features; and customer-facing systems, including real-time voice support, depend on frontier models being available, governed and affordable. The models these workloads need come from several vendors simultaneously, and the best model for a task can change on a timescale of weeks.

The obvious way to consume these models is for each application to integrate each vendor directly: hold the vendor's key, adopt its request format, handle its streaming semantics, and track its spend locally. That approach works for one team and one model. It does not survive an organization of thousands of engineers and hundreds of production workloads spanning several major providers. Direct integration fails along four compounding axes:
\begin{itemize}
\item \textbf{Credential sprawl.} Long-lived vendor keys proliferate across applications, with no central place to scope, rotate, attribute or revoke them, and no reliable link from a call to the human who authorized it.
\item \textbf{Format lock-in.} An application written against one vendor's API cannot adopt another vendor's model without code changes, so model choice, regional placement and failover become engineering projects rather than configuration.
\item \textbf{Ungoverned spend.} Cost accrues per key and per vendor, with no organization-aware budgets, no early warning and no lever for business-unit leadership.
\item \textbf{Uneven governance and visibility.} Each integration authenticates, logs, rate-limits and protects data differently; security and finance have no single plane on which to observe or control AI traffic.
\end{itemize}
This paper describes how we addressed these failure modes behind a single layer, the Cosmos LLM Gateway. Beyond a unified API and data plane, the system comprises a full governance control plane, a policy-as-code decision engine, a platform-identity service attested by corporate single sign-on (SSO), a durable usage-attribution pipeline and inline data-loss prevention (DLP). This paper documents the system as it now operates.

\para{Contributions.} We make the following contributions:
\begin{enumerate}
\item A \textbf{two-tier Envoy architecture} that runs I/O-bound governance out of process and streaming translation in process, joined by an explicit, contract-tested handoff (section~\ref{sec:dataplane}).
\item \textbf{Models, routing and pricing as validated data}: a registry whose loader enforces a \emph{serviceability invariant}, and a control-plane onboarding lifecycle that takes a model from staged to generally available without a deployment (sections~\ref{sec:registry} and~\ref{sec:onboarding}).
\item A \textbf{canonical intermediate-representation (IR) translation layer} that needs one codec per wire format instead of one converter per format pair, with fast paths that are lossless for client content and gateway-side prompt-cache shaping (section~\ref{sec:translation}).
\item A \textbf{layered access and spend policy model}: organization-anchored model restrictions, business-unit controls, and budgets whose scopes are views over per-model counters keyed by immutable model identifiers, which makes them retroactive within a window and migration-free. Policies are evaluated by a pure decision function separated from side effects (section~\ref{sec:policy}).
\item A \textbf{governed policy lifecycle} for platform and business-unit administrators: approvals, scheduled activation, pilots with tier ladders, shadow-to-authoritative rollout, threshold notifications and versioned bundle distribution with last-known-good semantics (section~\ref{sec:control}).
\item A \textbf{durable, priced, per-request usage record} that drives chargeback and analytics, alongside the per-model counters that drive budgets (section~\ref{sec:telemetry}), together with \textbf{production evidence, an overhead benchmark} and the lessons we would carry to similar systems (sections~\ref{sec:eval} and~\ref{sec:lessons}).
\end{enumerate}
The gateway governs which \emph{models} an AI client may call. A companion platform, the Cosmos.AI MCP Hub~\cite{mcphub}, governs which \emph{tools} it may reach over the Model Context Protocol~\cite{mcp}. Together they form the access and control plane for enterprise AI at PayPal.

\section{Requirements}
We derived the requirements from the failure modes above and from the operating constraints of a regulated financial-services company.

\para{R1 One API, approved models.} Clients use APIs they already know (OpenAI Chat Completions, Responses and Embeddings, Anthropic Messages, and Gemini generateContent) and reach approved models through the conversion paths the registry declares (table~\ref{tab:conv}). Changing the vendor or region behind a model is a gateway change, not an application change.

\para{R2 Brokered credentials.} Clients present only an internal credential: a personal access token for developers, an application credential bound to a registered application for business applications, or a platform-minted token (section~\ref{sec:identity}). The gateway obtains the upstream vendor credential, attributes its use to the calling user or application, and never returns it to the client.

\para{R3 Identity-aware policy.} Every request made on behalf of a person is checked against the caller's current position in the organizational hierarchy, so model restrictions and budgets follow people through reorganizations.

\para{R4 Configuration, not deployment.} Onboarding a model, changing a price, or adding a budget, restriction or notification is a validated data change.

\para{R5 Streaming-safe performance.} Translation and metering must not buffer streamed responses or add material time-to-first-token. Gateway overhead must be small and stable under load.

\para{R6 Data protection.} Where a use case's data classification requires it, the gateway applies detection to prompts and blocks, redacts or pseudonymizes detected personal and payment data before it reaches an external provider, with policy set per use case.

\para{R7 Explicit failure semantics.} Access control and data protection fail closed; observability fails open; an invalid configuration prevents startup instead of serving incorrectly.

\para{R8 Connections, not only requests.} Long-lived real-time sessions (voice calls of up to 30 minutes) must survive deployments, autoscaling and node upgrades.

\para{R9 Attribution by default.} Every governed request is designed to produce a durable, priced usage record attributable to its calling principal (a user or a registered application) and, where present, to a platform and workflow, with no application instrumentation.

\para{R10 No big-bang migration.} Hundreds of applications were already in production. Any plan that required every team to update its integration was not acceptable.

\section{Background and Related Work}
\para{Programmable proxies.} Envoy~\cite{envoy} is a widely deployed layer-7 (L7) proxy with mature TLS, HTTP/2, WebSocket, load-balancing and telemetry support. It offers two extension mechanisms that matter here. The \emph{external processing} filter (ext\_proc)~\cite{extproc} streams request and response phases to a gRPC service that may inspect and mutate headers, bodies and trailers and may return an immediate response; it runs out of process and can be configured to fail open or closed. \emph{Proxy-Wasm}~\cite{proxywasm} runs sandboxed WebAssembly modules inside the proxy's worker threads, which suits per-byte work on streaming bodies.

\para{LLM gateways.} Open-source LLM proxies provide a unified API across vendors, typically as an application-level service~\cite{portkey}. Agent Router (formerly Envoy AI Gateway)~\cite{envoyai} and agentgateway~\cite{agentgateway} build AI-specific routing on top of proxy data planes, and cloud providers now offer managed AI gateways~\cite{cfaigw,azureapim}. The closest design to ours, Agent Router, performs request translation, upstream authorization and token metering in an external processor, and names Envoy dynamic modules as a possible higher-performance alternative~\cite{envoyai}. A single streamed response can carry thousands of chunks, and we did not want to pay a network round trip to an external process for each of them. We therefore keep only I/O-bound decisions out of process and run per-chunk translation and metering in-proxy as WebAssembly (section~\ref{sec:dataplane}). We evaluated an off-the-shelf proxy before building. Basic routing worked, but at the time of evaluation it did not meet our requirements for real-time (WebSocket) traffic, identity-aware authorization, enforceable user- and organization-level budgets, or workload-identity-based cloud credentials. Our design shares the proxy-native approach of these projects. It differs mainly in the governance layer: organization-anchored policy, retroactive spend scopes, and a managed policy lifecycle.

\para{Authorization and policy as code.} Zanzibar~\cite{zanzibar} established relationship-based authorization at global scale. Open Policy Agent (OPA)~\cite{opa} popularized declarative, side-effect-free policy evaluated against injected data. We use OPA as the decision engine. The gateway gathers state and performs effects, which keeps decisions pure, testable and replayable (section~\ref{sec:policy}).

\para{Cost-aware model use and serving.} FrugalGPT~\cite{frugalgpt} and RouteLLM~\cite{routellm} route queries across models to trade cost against quality. Serving systems such as vLLM~\cite{vllm} and SGLang~\cite{sglang} optimize inference itself. These lines of work complement ours: a gateway is the natural place to enforce the budget decisions that a router optimizes for, and it sits above the serving systems that host open-weight models.

\begin{figure*}[t]
\centering
\resizebox{\textwidth}{!}{%
\begin{tikzpicture}[font=\small]
\fill[gray!6, rounded corners=8pt] (-0.45,1.95) -- (10.9,1.95) -- (10.9,-3.15) -- (15.7,-3.15) -- (15.7,1.95) -- (21.7,1.95) -- (21.7,-7.15) -- (-0.45,-7.15) -- cycle;
\draw[gray!60, dashed, line width=0.6pt, rounded corners=8pt] (-0.45,1.95) -- (10.9,1.95) -- (10.9,-3.15) -- (15.7,-3.15) -- (15.7,1.95) -- (21.7,1.95) -- (21.7,-7.15) -- (-0.45,-7.15) -- cycle;
\node[font=\scriptsize\itshape\bfseries, text=black!55, anchor=north west] at (-0.3,1.9) {PayPal private network};
\node[font=\scriptsize\itshape\bfseries, text=black!70] at (1.6,1.15) {Clients};
\node[font=\scriptsize\itshape, text=black!60] at (1.6,0.85) {OpenAI / Anthropic / Gemini SDKs};
\node[gbox, minimum width=3.4cm, minimum height=0.55cm] (c1) at (1.6,0.35) {Coding agents \& CLIs};
\node[gbox, minimum width=3.4cm, minimum height=0.55cm] (c2) at (1.6,-0.95) {Agentic workflows};
\node[gbox, minimum width=3.4cm, minimum height=0.55cm] (c3) at (1.6,-2.25) {Business applications};
\node[gbox, minimum width=3.4cm, minimum height=0.55cm] (c4) at (1.6,-3.5) {Real-time voice};
\def\ex{7.4}
\node[bbox, minimum width=4.2cm, minimum height=0.85cm] (tls) at (\ex,0.35) {TLS listener + path routing};
\node[obox, minimum width=4.2cm, minimum height=1.05cm] (ctl) at (\ex,-1.0) {control ext\_proc\\[-1pt]{\scriptsize auth $\cdot$ policy $\cdot$ DLP $\cdot$ credentials}\\[-2pt]{\scriptsize fail-closed}};
\node[wbox, minimum width=4.2cm, minimum height=0.85cm] (met) at (\ex,-2.35) {metrics ext\_proc\\[-1pt]{\scriptsize observability mode $\cdot$ fail-open}};
\node[ebox, minimum width=4.2cm, minimum height=0.85cm] (wasm) at (\ex,-3.55) {translate.wasm (Rust)\\[-1pt]{\scriptsize in-proxy translation $\cdot$ metering}};
\node[wbox, minimum width=4.2cm, minimum height=0.85cm] (rt) at (\ex,-4.8) {router + upstream clusters};
\draw[arr] (tls) -- (ctl); \draw[arr] (ctl) -- (met); \draw[arr] (met) -- (wasm); \draw[arr] (wasm) -- (rt);
\node[draw=cblue, dashed, rounded corners=6pt, line width=0.7pt, inner sep=6pt, fit=(tls)(rt)] (envoy) {};
\node[font=\small\bfseries, text=cblue!80!black, above=1pt of envoy.north] {Envoy data plane (autoscaled)};
\draw[line width=0.5pt, black!75] (c1.east) -- (4.2,0.35);
\draw[line width=0.5pt, black!75] (c2.east) -- (4.2,-0.95);
\draw[line width=0.5pt, black!75] (c3.east) -- (4.2,-2.25);
\draw[line width=0.5pt, black!75] (c4.east) -- (4.2,-3.5);
\draw[line width=0.5pt, black!75] (4.2,0.35) -- (4.2,-3.5);
\draw[arr] (4.2,0.35) -- (tls.west);
\def\px{13.3}
\node[font=\scriptsize\itshape\bfseries, text=black!70] at (\px,1.05) {External model providers};
\node[gbox, minimum width=4.2cm, minimum height=0.85cm] (p1) at (\px,0.35) {{\footnotesize Gemini Enterprise Agent Platform}\\[-1pt]{\scriptsize Gemini $\cdot$ Claude (global endpoint)}};
\node[gbox, minimum width=4.2cm, minimum height=0.55cm] (p2) at (\px,-0.85) {OpenAI};
\node[gbox, minimum width=4.2cm, minimum height=0.55cm] (p3) at (\px,-1.75) {xAI};
\node[gbox, minimum width=4.2cm, minimum height=0.55cm] (p5) at (\px,-2.65) {Azure-hosted models};
\node[ebox, minimum width=4.2cm, minimum height=0.85cm] (p4) at (\px,-4.8) {Self-hosted open-weight (GKE)\\[-1pt]{\scriptsize vLLM $\cdot$ SGLang $\cdot$ NVIDIA Dynamo}};
\draw[line width=0.5pt, black!75] (rt.east) -- (10.4,-4.8);
\draw[line width=0.5pt, black!75] (10.4,-4.8) -- (10.4,0.35);
\draw[arr] (10.4,0.35) -- (p1.west);
\draw[arr] (10.4,-0.85) -- (p2.west);
\draw[arr] (10.4,-1.75) -- (p3.west);
\draw[arr] (10.4,-2.65) -- (p5.west);
\draw[arr] (10.4,-4.8) -- (p4.west);
\def\cx{19.2}
\node[obox, minimum width=4.0cm, minimum height=0.85cm] (m1) at (\cx,0.35) {Management API\\[-1pt]{\scriptsize platform \& BU admins}};
\node[obox, minimum width=4.0cm, minimum height=0.85cm] (m2) at (\cx,-1.0) {Registry, prices \& policy\\[-1pt]{\scriptsize durable system of record}};
\node[obox, minimum width=4.0cm, minimum height=0.85cm] (m3) at (\cx,-2.35) {Projector + bundle publisher\\[-1pt]{\scriptsize versioned, leader-elected}};
\node[obox, minimum width=4.0cm, minimum height=0.85cm] (m4) at (\cx,-3.55) {Platform token mint\\[-1pt]{\scriptsize SSO $\rightarrow$ signed JWT}};
\node[draw=corange, dashed, rounded corners=6pt, line width=0.7pt, inner sep=6pt, fit=(m1)(m4)] (cp) {};
\node[font=\small\bfseries, text=corange!90!black, above=1pt of cp.north] {Control \& identity planes};
\draw[darr] (m3.west) -- (16.4,-2.35) -- (16.4,-3.75) -- (10.0,-3.75) -- (10.0,-1.0) -- (ctl.east);
\node[font=\scriptsize\itshape, text=black!60] at (13.3,-3.5) {policy bundles};
\node[ebox, minimum width=9.6cm, minimum height=0.5cm] (tel) at (6.4,-6.6) {Telemetry plane: metrics $\cdot$ traces $\cdot$ per-request usage records $\rightarrow$ warehouse};
\draw[darr] (envoy.south) -- (\ex,-6.33);
\end{tikzpicture}}
\caption{Architecture. One governed entry point fronts enterprise LLM traffic. On the unified API, the control service makes I/O-bound decisions out of process and fails closed, and the Wasm filter translates and meters in process. Real-time sessions use a connection-oriented path tuned for long-lived streams. The control plane publishes versioned policy bundles that the data plane enforces on every request. The shaded region is PayPal's private network: self-hosted open-weight models run inside it, and external providers are reached with brokered credentials.}
\label{fig:arch}
\end{figure*}
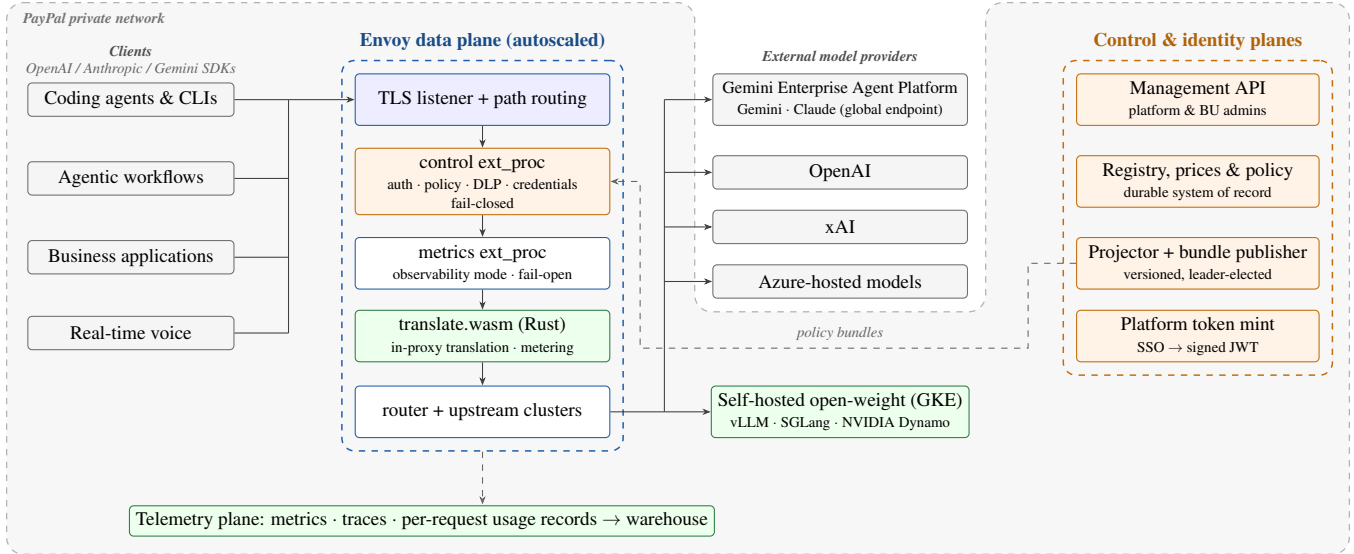

\section{System Overview}
Figure~\ref{fig:arch} shows the system. A single TLS listener terminates all traffic and routes it by path through the same Envoy pod. Unified API requests pass through the control path (authentication, policy, data protection and credential brokering), translation and metering in the Wasm filter, and an upstream call with a brokered credential (section~\ref{sec:dataplane}). Real-time voice APIs run over WebSocket; for these, Envoy holds each long-lived connection on its event loop and forwards audio frames as they arrive, without body buffering (section~\ref{sec:reliability}).

Behind the data plane sit three supporting planes. The \emph{control plane} holds the authoritative model registry, prices and governance policy, exposes a management API for platform and business-unit (BU) administrators, and publishes versioned policy bundles. The \emph{identity plane} mints verifiable JSON Web Tokens (JWTs) for platforms that call the gateway on behalf of users. The \emph{telemetry plane} carries metrics, traces and durable per-request usage records into the analytics warehouse.

\para{Deployment.} The gateway runs on Google Kubernetes Engine (GKE) in multiple regions with disaster recovery. Closed-weight models such as Gemini and Claude are served, where supported, through the global endpoint of Google's Gemini Enterprise Agent Platform (formerly Vertex AI Platform), which avoids dependence on a single regional endpoint. Open-weight models run on an internal GKE inference stack built on vLLM~\cite{vllm}, SGLang~\cite{sglang} and NVIDIA Dynamo~\cite{dynamo}. Other providers are reached over the network with brokered credentials. The caller does not need to know where a model runs.

\section{Data Plane: Two-Tier Extensibility}\label{sec:dataplane}
\subsection{Splitting work by latency profile}
The most consequential design decision was to refuse the trade-off between \emph{governed} and \emph{fast}. Governance decisions are I/O-bound: they need identity validation, the organizational hierarchy, spend counters and secret material. They therefore run out of process in a Go ext\_proc service that scales independently of the proxy. Payload work is CPU-bound, sits on the critical path of every streamed byte, and needs no network I/O. It therefore runs inside Envoy as a Rust Proxy-Wasm filter, so that a streamed response is translated and metered chunk by chunk without waiting on any out-of-process call.

\begin{table}[t]
\centering\footnotesize
\begin{tabular}{@{}lll@{}}
\toprule
\textbf{Setting} & \textbf{control ext\_proc} & \textbf{metrics ext\_proc}\\
\midrule
Failure mode & fail closed & fail open\\
Request headers & send & send\\
Request body & buffered & streamed (mirrored)\\
Response headers & skip & send\\
Response body & none & streamed (mirrored)\\
Response trailers & send & send\\
Blocking & yes & no (observability mode)\\
Dynamic metadata & attribution, cost & --\\
\bottomrule
\end{tabular}
\caption{Default ext\_proc processing modes. Response bodies never wait on an out-of-process call unless a use case's data-loss-prevention policy requires response handling (section~\ref{sec:identity}).}
\label{tab:extproc}
\end{table}

Below, the \emph{control service} is the Go service behind the control ext\_proc, the \emph{metrics service} is the service behind the metrics ext\_proc, and the \emph{Wasm filter} is the in-process translation filter (translate.wasm in fig.~\ref{fig:arch}). The HTTP filter chain runs in a fixed order: (1) the control ext\_proc, configured to fail closed, which receives the buffered request body because the routing key (\texttt{model}) is in the body; (2) a metrics ext\_proc, configured to fail open, which runs after control and therefore observes the request as authorized and rewritten; (3) the translation Wasm filter; (4) a small buffer filter that restores \texttt{Content-Length} after translation, because some upstreams reject chunked request bodies and Proxy-Wasm cannot set headers during the body phase; and (5) the router. Table~\ref{tab:extproc} lists the ext\_proc processing modes. The metrics service runs in Envoy's \emph{observability mode}, in which bodies are mirrored to it asynchronously and never held. The control service buffers only the request body, skips response headers and bodies entirely, and is invoked again only for response trailers, where it reads usage. By default, the streamed response body therefore sits on the blocking critical path of exactly one extension, the in-process filter; the metrics service sees a non-blocking mirror, and the control service is consulted again only for the trailers at the end of the stream. In production, the metrics service derives metrics from mirrored bodies without persisting their content.

\subsection{Request lifecycle}
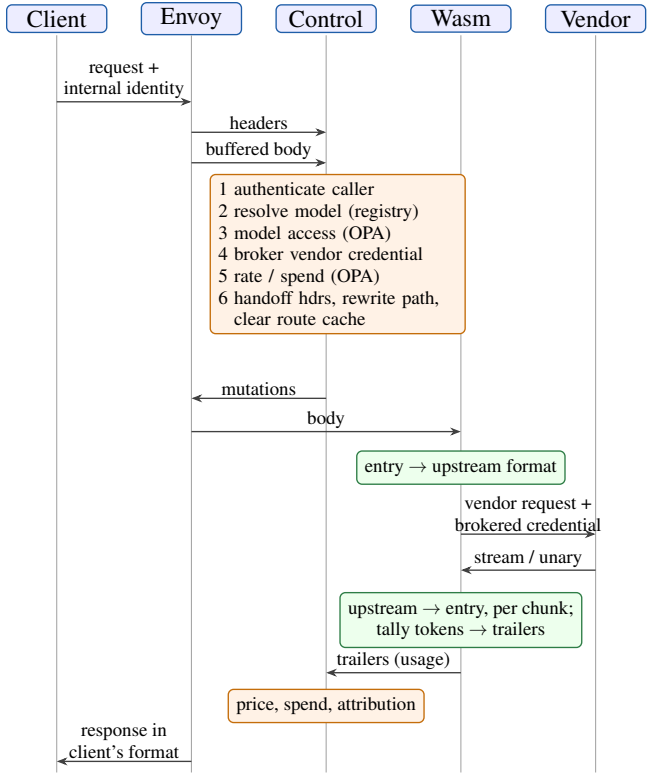
\begin{figure}[t]
\centering
\resizebox{\columnwidth}{!}{%
\begin{tikzpicture}[font=\small, y=0.36cm]
\foreach \n/\x in {Client/0, Envoy/1.75, Control/3.5, Wasm/5.25, Vendor/7.0} {
  \node[bbox, minimum width=1.3cm, inner sep=2pt] (h\n) at (\x,0) {\n};
  \draw[black!35, line width=0.4pt] (\x,-0.6) -- (\x,-27.2);
}
\def\msg#1#2#3#4{\draw[arr] (#1,#3) -- node[above, font=\scriptsize, align=center, inner sep=1pt]{#4} (#2,#3);}
\msg{0}{1.75}{-3.0}{request +\\internal identity}
\msg{1.75}{3.5}{-4.1}{headers}
\msg{1.75}{3.5}{-5.2}{buffered body}
\node[obox, align=left, font=\scriptsize, text width=3.1cm, anchor=north] at (3.65,-5.6) {1 authenticate caller\\2 resolve model (registry)\\3 model access (OPA)\\4 broker vendor credential\\5 rate / spend (OPA)\\6 handoff hdrs, rewrite path,\\\phantom{6 }clear route cache};
\msg{3.5}{1.75}{-13.7}{mutations}
\msg{1.75}{5.25}{-14.9}{body}
\node[ebox, font=\scriptsize] at (5.25,-16.2) {entry $\rightarrow$ upstream format};
\msg{5.25}{7.0}{-18.6}{vendor request +\\brokered credential}
\msg{7.0}{5.25}{-19.9}{stream / unary}
\node[ebox, font=\scriptsize, align=center] at (5.25,-21.7) {upstream $\rightarrow$ entry, per chunk;\\tally tokens $\rightarrow$ trailers};
\msg{5.25}{3.5}{-23.6}{trailers (usage)}
\node[obox, font=\scriptsize] at (3.5,-24.8) {price, spend, attribution};
\msg{1.75}{0}{-26.8}{response in\\client's format}
\end{tikzpicture}}
\caption{Lifecycle of a unified request. Routing decisions happen in the body phase because the model is in the body. Streamed response bytes pass through only the in-process filter on the critical path.}
\label{fig:lifecycle}
\end{figure}

Figure~\ref{fig:lifecycle} traces a unified request. In the \emph{request-headers} phase the control service records the entry wire format (from the path) and the caller's credential; Envoy then delivers the buffered, size-bounded body. In the \emph{request-body} phase it authenticates the caller, extracts the model, resolves it through the registry, evaluates model access, brokers the upstream credential, evaluates rate and spend policy, sets the handoff headers for the Wasm filter, rewrites the path, and asks Envoy to clear its route cache so that Envoy re-selects the vendor route and cluster for the rewritten request. It also attaches attribution metadata for the access log. The Wasm filter then translates the request into the upstream format. On the response, the filter translates each chunk back into the client's format, taps token usage from the stream, and emits usage as response trailers. In the \emph{response-trailers} phase the control service prices the request from the registry, records spend, attaches cost metadata, and invalidates a cached upstream credential if the vendor returned 401 or 403.

\subsection{A contract between two languages}
The Go control service and the Rust filter communicate only through a small, versioned contract of gateway-internal request metadata and response trailers, which carry the entry and upstream wire formats, the upstream model, token usage, a count of unclassified fields and the tools the model invoked. The contract lives in one machine-readable file that also names the access-log metadata fields (section~\ref{sec:telemetry}). Test suites in both languages assert their constants against it, so any drift fails the build, and an unrecognized value is never reinterpreted. In practice, this explicit cross-language contract removed a class of integration bugs that otherwise surfaced only in end-to-end tests.

\subsection{Failure semantics}
Each component on the request path had to answer one question: \emph{if this breaks, what happens to the request?} Table~\ref{tab:failure} summarizes the answers. The asymmetry is deliberate. An access or data-protection rule that silently stops enforcing during an outage is not a rule, so those decisions fail closed. A telemetry outage must never become a traffic outage, so observability fails open. New budgets are rolled out in shadow mode first, which is designed to keep a defect in a newly introduced policy from taking AI tooling offline for the whole organization. Together these rules make degradation graceful and predictable: telemetry failures never reach the caller, a failed policy update leaves the last good revision in force, and an invalid configuration is rejected before it serves traffic.

\section{Models, Routing and Pricing as Validated Data}\label{sec:registry}
Early versions encoded vendor selection, upstream paths and capability flags in code. That made every new model a deployment, and nothing could be edited by an administrator. We replaced it with a \emph{registry}: a typed, declarative model of everything the gateway supports. Model resolution, upstream paths, capability gating and prices all come from the registry. The code keeps a path-template substitutor, the credential resolvers (which perform I/O), and a small number of provider-specific adaptations for vendor quirks, such as session-affinity headers.

\begin{table}[t]
\centering\footnotesize
\begin{tabular}{@{}p{0.47\columnwidth}p{0.47\columnwidth}@{}}
\toprule
\textbf{Failure} & \textbf{Behavior}\\
\midrule
Control service unavailable & Request rejected (fail closed)\\
Access-policy evaluation error & Deny\\
Spend-policy evaluation error & Deny when enforcing; logged in shadow mode\\
Data-protection check cannot complete & Request not sent (fail closed)\\
Invalid registry / policy at boot & Process refuses to start\\
Policy bundle reload fails & Keep last-known-good revision\\
Metrics service unavailable & Traffic unaffected (fail open)\\
Client reset mid-stream & Usage record still written (section~\ref{sec:telemetry})\\
\bottomrule
\end{tabular}
\caption{Failure semantics on the request path.}
\label{tab:failure}
\end{table}

\para{Entities.} The registry defines \emph{brands} (who makes a model), \emph{API families} (chat, responses, embeddings), \emph{wire formats} (e.g., Anthropic Messages, OpenAI Chat, OpenAI Responses, Anthropic-on-Vertex, Gemini generateContent), \emph{credentials} (how a provider authenticates, holding secret \emph{names} and never values), \emph{providers}, \emph{clusters}, \emph{provider operations} keyed by (provider, deployment format, API family) that map to a path template and streaming mode, \emph{models} (with aliases, tags, entry formats and access metadata), \emph{deployments} (physical serving locations with region, priority and price), \emph{entry routes} and \emph{converters} (legal translation pairs).

\para{Resolution.} For a request with entry path $p$ and model identifier $x$, the data plane computes:
\noindent\begin{minipage}{\columnwidth}
\begin{lstlisting}[style=code,language=pseudo]
entry  = entry_routes[p].format
model  = resolve(x)             # canonical id or alias
family = model.upstream_family or family(entry)
dep    = model.deployment_for(entry)
op     = operations[(dep.provider, dep.format, family)]
up_fmt = canonical(family) or dep.format
require converter(entry -> up_fmt)   # identity or declared
path   = substitute(op.path, dep)    # project, region, model, verb
cred   = providers[dep.provider].credential
\end{lstlisting}
\end{minipage}
Capability gating is therefore data. If no responses-family operation exists for a provider, that provider does not support the Responses API, and no code needs to check for it. An optional per-model \texttt{upstream\_family} pins the vendor surface when the client's family would choose the wrong one. For example, a coding agent may speak a chat-family API while the vendor requires its responses endpoint for tool use combined with reasoning.

\para{Pricing as configuration.} Each deployment carries its price in USD per million tokens for input, output, cache reads, cache writes and extended-lifetime cache writes. A price may also define a \emph{long-context tier} that applies when the prompt (input, cache-read and cache-write tokens) exceeds a threshold. For a request with token counts $n_k$ and the applicable tier's rates $p_k$, the cost in micro-dollars is $\sum_k n_k p_k$. A price change is therefore a registry edit that takes effect within one publication cycle, and the management API validates prices on every write: units come from an allow list, every rate is non-negative, and unknown keys are rejected.

\para{Fail-closed validation and the serviceability invariant.} The loader turns raw configuration into a typed registry or returns an aggregated error that aborts startup. It checks that every reference resolves, that credential kinds match their providers, that operations and aliases do not collide, and that the following invariant holds:
\begin{multline*}
\forall\, m \in \mathscr{M},\ \forall\, f \in \mathrm{entry}(m) : \exists\, d \in \mathrm{dep}(m) :\\
\mathrm{op}(d,f) \neq \bot \ \wedge\ \mathrm{conv}\big(f, \mathrm{up}(d,f)\big).
\end{multline*}
In words, every format a model advertises is backed by at least one deployment that has both an operation and a converter. The model-discovery endpoint is rendered from the same registry, so discovery and routing cannot disagree. Appendix~\ref{app:registry} gives an abridged registry excerpt.

\para{Admission and discovery.} Model discovery is rendered from the registry, and its visibility is configurable per model. The caller is authenticated before the model is looked up, and unknown and policy-denied models return the same response (HTTP 404, ``model not found or not available''), so the API does not reveal whether a hidden or limited-access model exists. The specific reason is recorded server-side, in logs and audit reason codes, for operators.

\para{From file to database.} The registry began as a version-controlled file and now has a durable store behind the management API (section~\ref{sec:onboarding}). A leader-elected projector rebuilds the served registry deterministically whenever the store's epoch changes and publishes it through the same bundle channel as policy (section~\ref{sec:control}). The data plane validates the projected registry when it loads it; an invalid or missing bundle fails configuration. Registry and policy epochs are independent, so a catalog change never requires rewriting policy rules, and vice versa; registry and access data are published in the same bundle.

\begin{table}[t]
\centering\footnotesize
\setlength{\tabcolsep}{2pt}
\begin{tabular}{@{}lcccc@{}}
\toprule
 & \multicolumn{4}{c}{\textbf{Upstream format}}\\
\cmidrule(l){2-5}
\textbf{Entry format} & \textbf{OAI Chat} & \textbf{OAI Resp.} & \textbf{Anth.@Vertex} & \textbf{Gemini}\\
\midrule
Anthropic Messages & T & T & R & T\\
OpenAI Chat & T$^{*}$ & T & T & T\\
OpenAI Responses & -- & I & -- & --\\
Gemini generateContent & -- & -- & -- & I\\
\bottomrule
\end{tabular}
\caption{Conversion plan by entry and upstream format: OAI = OpenAI. I = identity, R = raw envelope rewrite, T = translation through the IR, -- = not offered (by registry configuration, not code). \emph{Vertex} denotes the Gemini Enterprise Agent Platform API. $^{*}$Same-format round trip that preserves Gemini tool-call thought signatures (section~\ref{sec:translation}).}
\label{tab:conv}
\end{table}

\section{In-Proxy Translation}\label{sec:translation}
\para{A canonical IR: $N$ codecs, not $N(N-1)$ converters.} Every wire format decodes into, and encodes out of, one canonical intermediate representation (IR). An any-to-any conversion is \emph{decode} $\rightarrow$ \emph{IR} $\rightarrow$ \emph{encode}. Each codec translates only between the IR and one wire format, with no knowledge of other formats, so adding a format costs one codec rather than one converter per existing format. The system has five general-purpose codecs, plus raw adapters for a small number of internal formats.

\para{Three conversion plans.} Building the IR is necessary only when the envelope cannot simply be rewritten. The converter chooses among:
\begin{itemize}
\item \textbf{Identity}: source equals target, and bytes pass through unchanged.
\item \textbf{Raw}: same family with a different envelope (e.g., Anthropic Messages to Anthropic-on-Vertex). The message content is kept verbatim and only envelope fields are rewritten (plus the gateway's cache breakpoints), which preserves vendor-only fields such as cache control and extended thinking that an IR round-trip might drop.
\item \textbf{Translate}: a full decode--IR--encode, for example from Anthropic Messages to OpenAI Chat.
\end{itemize}
Both fast paths are lossless for client content because they never build the IR. Table~\ref{tab:conv} shows the plan for the main entry and upstream pairs. OpenAI Chat to OpenAI Chat is a deliberate exception to identity. Gemini, served through an OpenAI-compatible chat endpoint, returns a \texttt{thought\_signature}~\cite{geminisig} on each tool call (in a vendor extension field, \texttt{extra\_content}) and requires it to be sent back on the next turn; some agent clients drop unrecognized fields when they echo the conversation, which would break multi-turn tool use. On the response, the codec therefore reversibly encodes the signature into the tool-call identifier, which clients must echo verbatim to correlate tool results. On the next request, it decodes the identifier, restores the original ID and reinstates the signature in the extension field before forwarding upstream. The mechanism is stateless and applies only to signatures attached to tool calls; the Anthropic Messages and OpenAI Responses codecs use the same identifier encoding.

\para{Streaming.} Server-sent event streams are converted incrementally. A decoder maps wire chunks to IR stream events (message start, content-block start, text and tool-argument deltas, block stop, usage, message stop), and an encoder maps those events to the target's event grammar. Per-stream decode and encode state carries partial bytes across chunk boundaries, maps the wire stream's content-block indices onto the IR's indexed blocks, and buffers nothing beyond an incomplete event. Reasoning signatures that some vendors require to be echoed on the next turn are carried through the IR, so multi-turn tool use survives a format change. Token usage is tapped from the stream and reported in trailers without altering the passthrough bytes. Because Proxy-Wasm forbids header writes during the body phase, all metadata produced while streaming leaves the filter as trailers.

\para{Schema drift is a signal, not silent loss.} Vendors evolve their APIs continuously. Request fields the codecs cannot classify are counted and surfaced as a trailer and a metric. In addition, a coverage test fails the build when a known field is unclassified, which forces a human decision instead of silent data loss.

\para{Gateway-side prompt caching.} Coding agents resend long, stable prefixes (system prompts, tool definitions, conversation history) on every turn. The filter injects cache breakpoints on the last tool definition, the system prompt and the most recent content block, within the vendor's limit of four breakpoints~\cite{anthropiccache} and without removing breakpoints the client already set, and it requests an extended (one-hour) cache lifetime for all of them. This shaping applies on the Anthropic-on-Vertex upstream path. For providers that support it, the control service can also add a stable, hashed session-affinity key derived from the caller's identity (and the client's session where available), enabled per provider or for a configurable share of traffic, so that repeated turns land on warm caches. Clients get these benefits without changing any code; the effect is reported in section~\ref{sec:eval}.

\section{Identity, Credentials and Data Protection}\label{sec:identity}
\para{Inbound credentials.} Every caller of the unified API presents an internal credential, never a vendor key. Developers use a personal access token tied to their corporate identity. Business applications present an application credential bound to a registered application, so that usage is attributed to that application rather than to an individual. Platforms that act on behalf of users present tokens from the mint service described below. Personal and platform-minted tokens resolve to a person, whom the policy layer evaluates against the organizational hierarchy; application credentials resolve to the registered application.

\para{Outbound: brokered, keyless credentials.} Clients of the unified API never hold a vendor key, so a leaked application configuration does not expose a vendor credential. The registry declares a credential \emph{kind} per provider and the gateway resolves it per request: per-user upstream credentials brokered by the gateway, OAuth2 client-credentials grants for directory-backed providers, the gateway's own workload identity for cloud-hosted models, and, for application traffic, a short-lived credential scoped to the calling application. Secrets are fetched at runtime from a managed secret store using the platform's own identity; secret values are not placed in code or static configuration, and request logging redacts credentials.

\para{SSO-attested platform identity.} Platforms that call the gateway on behalf of users, such as agent harnesses and internal developer platforms, need a trustworthy way to identify themselves and label workflows. We introduced a \emph{mint} service that converts ``a user signed in to platform $P$ through corporate SSO'' into a signed, time-bounded JSON Web Token (JWT)~\cite{jwt} naming the user, the platform and optional attribution tags (e.g., workflow and run). Its design choices are:
\begin{itemize}
\item \textbf{No shared platform secret.} A platform is identified by the OAuth client identifier bound to its OpenID Connect (OIDC)~\cite{oidc} flow. The mint service accepts only a verified \texttt{id\_token} whose authorized party (or, when absent, audience) resolves to a registered, active platform. The identity provider issues such tokens only through that client's flow.
\item \textbf{The user is the subject; the platform is an attribution dimension.} The platform cannot choose the subject. Spend is charged to the user's budget, and the platform appears on every usage record.
\item \textbf{Offline verification.} Tokens are signed, with keys published through a JSON Web Key Set (JWKS) endpoint~\cite{jwk}. The control service verifies them locally, with no per-request call to the mint service.
\item \textbf{Bounded lifetime.} Token lifetime is capped per platform and never exceeds the expiry of the presented identity token, so re-minting cannot extend access beyond what the identity provider issued.
\item \textbf{Tags are attribution, never authority.} Tag keys are allow-listed per platform, size-bounded and rejected when unknown. They never expand routing, authorization or quota.
\item \textbf{Authentication tokens do not persist.} The \texttt{id\_token} and the minted token are never logged or stored. Every decision, including rejections, is audited by reason code.
\end{itemize}
Minted tokens can also carry a \emph{workflow budget}, bounded at mint time and signed, that caps the total spend of a workflow across all of its users. It is enforced in addition to user budgets and can never bypass them.

\subsection{Data loss prevention}
\begin{figure}[t]
\centering
\resizebox{\columnwidth}{!}{%
\begin{tikzpicture}[font=\small]
\node[gbox, minimum width=2.0cm, minimum height=0.95cm] (rq) at (0,0) {Request\\[-2pt]{\scriptsize may contain personal data}};
\node[obox, minimum width=2.0cm, minimum height=0.95cm] (dt) at (2.55,0) {Detect\\[-2pt]{\scriptsize rules +}\\[-3pt]{\scriptsize internal model}};
\node[obox, minimum width=2.3cm, minimum height=0.95cm] (ap) at (5.25,0) {Apply policy\\[-2pt]{\scriptsize block $\cdot$ redact $\cdot$}\\[-3pt]{\scriptsize pseudonymize}};
\node[gbox, minimum width=1.6cm, minimum height=0.95cm] (mp) at (7.75,0) {Model\\[-2pt]provider};
\node[gbox, minimum width=2.0cm, minimum height=0.6cm] (rs) at (0,-1.5) {Response};
\node[ebox, minimum width=3.4cm, minimum height=0.6cm] (ro) at (5.0,-1.5) {Restore original values};
\draw[arr] (rq) -- (dt); \draw[arr] (dt) -- (ap); \draw[arr] (ap) -- (mp);
\draw[arr] (mp.south) |- (ro.east);
\draw[arr] (ro.west) -- (rs.east);
\end{tikzpicture}}
\caption{Data-loss-prevention flow. Detected sensitive content is blocked, redacted or replaced with format-preserving surrogates inside the enterprise boundary before the provider call; surrogates are mapped back in the response.}
\label{fig:dlp}
\end{figure}
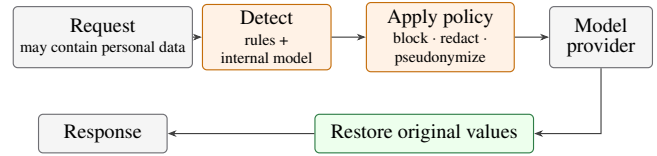

Prompts in a payments company can contain personal and payment data that should not reach an external model provider. For use cases whose data classification requires it, the gateway therefore applies a data-loss-prevention (DLP) layer on the request path (fig.~\ref{fig:dlp}).

\para{Detection inside the enterprise boundary.} Detecting sensitive data in free text is harder than matching patterns: the same date or number can be harmless in one sentence and sensitive in the next. Detection therefore combines deterministic pattern rules with an internally hosted, purpose-trained detection model that uses the surrounding context. Both run inside the enterprise boundary, so content is inspected before any vendor call is made. Detection is probabilistic, and accuracy varies by entity type. DLP therefore reduces exposure rather than guaranteeing that no sensitive value is ever sent, and it complements, rather than replaces, use-case data classification and provider contractual controls.

\para{Policy-driven actions.} Each use case is configured with its own DLP policy, because data classification, security zone and tolerance for false positives differ between use cases. From most to least restrictive, a policy can \emph{block} a request that contains sensitive data; \emph{redact} the sensitive spans by masking or removing them; or \emph{pseudonymize} them, replacing each value with a realistic surrogate of the same format and restoring the original values in the response. For these use cases the gateway also handles the response body: the original values that were swapped out are held for the duration of the request and substituted back for their surrogates before the response is returned to the client. Redaction is simple but can distort the task (a model asked to apply an age-based rule cannot reason about a masked date). Pseudonymization preserves the format of each detected value while the provider never sees the detected originals, which makes it well suited to conversational use cases.

\para{Operational properties.} DLP is integrated into the gateway API and is enforceable per use case: an application adopts it through per-use-case configuration and a request-level enforcement flag rather than an extra service call or SDK change. Inspection fails closed: if a check cannot complete, the request is not sent. DLP events are logged for security monitoring and audit. Together with a mediated path, attribution and consistent usage records, these are technical controls that can contribute to the organization's broader compliance programs; the gateway does not by itself establish compliance with any regulatory framework.

\section{The Policy Model}\label{sec:policy}
Governance answers three independent questions: \emph{who may use which model}, \emph{how much may be spent}, and \emph{how fast}. The policies in this section govern person-attributed traffic: developers, and users acting through platforms. Application traffic is governed through its application registration and attributed by application identifier in the usage record (section~\ref{sec:telemetry}).

\subsection{Model restrictions and access layers}
Access is layered (table~\ref{tab:access}). Each access layer can only narrow what the layers above it allow (an authoritative pilot's cohort lists decide access for its enrolled users, section~\ref{sec:bupolicies}), and each is evaluated before any vendor credential is issued. Spend caps work differently: an approved business-unit override can raise a platform cap, but only within the platform's locks (sections~\ref{sec:spend} and~\ref{sec:bupolicies}). Model restrictions name organization \emph{leaders} rather than teams: a model may carry a leader deny list, an optional leader allow list, and explicit per-person exemptions. A caller belongs to a leader's organization when that leader appears in the caller's upward reporting chain. The chain is resolved from the people directory through a cache, with the organization snapshot distributed in the policy bundle as a fallback (section~\ref{sec:control}). Because rules name leaders rather than team lists, a reorganization that moves a team automatically moves its access and budgets, with no policy edits.

\begin{table}[t]
\centering\footnotesize
\begin{tabular}{@{}p{0.27\columnwidth}p{0.25\columnwidth}p{0.38\columnwidth}@{}}
\toprule
\textbf{Layer} & \textbf{Set by} & \textbf{Effect}\\
\midrule
Catalog admission & platform (registry) & unknown model $\rightarrow$ 404\\
Model restriction & platform admin & leader allow/deny lists and exemptions $\rightarrow$ 404 (same as unknown)\\
BU allow/block list & platform admin & a model is open to, or closed to, named business units\\
BU model toggle & BU admin & switch a model off for one's own BU; cannot override a platform lock\\
Limited access & platform admin & bounded named roster with its own spend controls and expiry\\
Pilot cohort & platform admin & date-bounded allow/restrict lists per cohort\\
\bottomrule
\end{tabular}
\caption{Access-control layers. Each layer narrows the access granted by the one above it, except authoritative pilot cohorts (section~\ref{sec:bupolicies}); spend caps are separate and can be raised within platform locks (section~\ref{sec:spend}).}
\label{tab:access}
\end{table}

\subsection{Spend: scopes are views, not storage}\label{sec:spend}
Let $\mathscr{M}$ be the set of canonical model identifiers and $\tau : \mathscr{M} \rightarrow 2^{\mathscr{T}}$ assign each model a set of tags from $\mathscr{T}$, the set of registry tags, along several axes: everything, vendor, model family, license class (proprietary or open-weight), or any operator-defined axis. A \emph{scope} $s$ is a tag or a single model, and resolves to $\mathscr{M}_s = \{m : s \in \tau(m)\} \cup (\{s\} \cap \mathscr{M})$. For budgeting, spend for user budgets is accumulated \emph{only} as counters $c(u,m,\omega)$ in micro-dollars, keyed by user $u$, immutable model identifier $m$ and window instance $\omega$ (daily, weekly or monthly, on a single fixed calendar timezone for all regions). Here $u$ ranges over people; application traffic is attributed by application identifier (section~\ref{sec:telemetry}), and workflow budgets (section~\ref{sec:identity}) are accumulated in counters of their own.

A budget rule consists of one or more \emph{entries} $e = (P_e, s_e, w_e, \theta_e, A_e)$, each with a population predicate $P_e$ (user tiers, organization leaders, business unit, cohort and exemptions), a scope, a window, a threshold and a set of actions. Rules carry a precedence weight, and entries are ordered by it. The utilization of entry $e$ for user $u$ at time $t$ is
\[
\rho_e(u,t) = \frac{1}{\theta_e} \sum_{m \in \mathscr{M}_{s_e}} c\big(u, m, \omega_{w_e}(t)\big).
\]
For a request by $u$ for model $m^{*}$, the request is \emph{blocked} by the first entry, in precedence order, that has a block action, satisfies $P_e(u) \wedge \rho_e(u,t) \geq 1$, covers the requested model ($m^{*} \in \mathscr{M}_{s_e}$), and is not \emph{superseded}. Otherwise it is \emph{paused} by the first matching entry with an active pause that covers $m^{*}$. Scoping blocks to the requested model means a cap on one model family stops only that family: a user who exhausts a frontier-model budget can keep working on other models. A platform entry is superseded for a user when an approved business-unit override for that user's BU covers the same window and scope with a \emph{higher} threshold, so an approved BU increase outranks the platform baseline for that BU only. Separately, the \emph{governing} limit reported to the user is the block or pause entry with the highest utilization.

Actions compose. \emph{Block} is stateless and recomputed per request, so it clears automatically when the window rolls over. \emph{Pause} blocks for a fixed duration and fires at most once per rule and window, so a user who returns hours later is not penalized again. \emph{Message} adds a live status line, and \emph{email} notifies the user or, as a deferred digest, the leadership chain (section~\ref{sec:notify}). A separate \emph{no-promote} action can bar a population from self-service tier promotion (section~\ref{sec:tiers}).

This formulation has a property that matters operationally. Because tags are resolved at evaluation time over counters keyed by immutable model identifiers, \textbf{a newly created tag or rule immediately governs spend recorded earlier in the current window, before the rule existed}. No backfill, key rewrite or migration is needed. A business-unit policy such as ``cap each person in this BU at a monthly amount for Opus-class models'' (a vendor's largest model family) is a single data change that takes effect over the month to date. Business-unit policies are entries whose population predicate selects that unit; their caps apply per person (section~\ref{sec:control}). Retroactive \emph{scope} is not retroactive \emph{repricing}: counters are already denominated in micro-dollars at the price in force when each request completed, so a later price change affects only subsequent requests. The same property lets a draft rule be dry-run against current usage before it is enabled (section~\ref{sec:control}). Appendix~\ref{app:rules} shows the rule language.

\para{Rate limits and workflow budgets.} Per-minute rate limits are available to protect shared upstream quotas. Workflow budgets (section~\ref{sec:identity}) apply in addition to user rules, and whichever limit trips first blocks.

\subsection{User tiers and self-service promotion}\label{sec:tiers}
Users belong to one of three tiers with progressively larger allowances, derived from enterprise entitlement roles. Promotion is self-service through an internal access workflow. Promotion is subject to usage-based eligibility, and promotions expire automatically. Any budget entry can also carry a no-promote gate, either unconditionally for a population or only while the requesting user's own spend in a scope is at or above a threshold, so leadership can pause promotions for an organization without editing user records.

\subsection{A pure decision function}\label{sec:pure}
We moved policy evaluation from imperative code to OPA~\cite{opa} and its policy language, Rego. The dividing line is strict: \textbf{the gateway gathers state and performs effects; Rego decides}. Before evaluation, the control service assembles an input document containing the caller, canonical model, tier, organization chain, the summed spend per (scope, window), pause state, window boundaries and the enforcement switch. Evaluation returns a decision (block or allow, the limiting entry, a retry-after hint and per-limit utilization). The gateway then performs any side effects: updating counters, recording pauses and one-shot notifications, and queuing emails. Listing~\ref{lst:rego} gives a simplified illustration.

\begin{figure}[t]
\begin{lstlisting}[style=code,language=rego,caption={Simplified, illustrative spend decision in Rego (not the production policy). Spend sums are supplied by the gateway; the policy only decides. In shadow mode nothing blocks, but utilization is still computed and reported.},label={lst:rego},captionpos=b]
package gateway.spend
import rego.v1
import data.gateway.util

# entries that apply to this caller (tier, org, BU,
# cohort, exemptions), with utilization in percent
applicable contains entry if {
  some e in data.gateway.entries
  util.population_matches(e, input)
  pct := util.pct(input.spent[e.scope_window], e.threshold)
  entry := object.union(e, {"pct": pct})
}

blocking := util.first_by_precedence([e |
  some e in applicable
  some a in e.actions
  a.type == "block"
  e.pct >= 100
  util.covers(e.scope, input.model)
  not util.superseded_by_bu_override(e)
])

decision := {"block": true, "limit": blocking.id} if {
  input.enforce
  blocking
} else := {"block": false}
\end{lstlisting}
\end{figure}

The policy is organized into isolated modules: access, spend, rate limiting, alerts, pilots, validation, shared utilities and read-only administrative queries. The pilot module is deliberately isolated from platform and BU policy documents, and the policy test target checks that separation. On the request path, an enrolled pilot decision overrides the platform and BU decision only when pilot enforcement is authoritative.

Keeping decisions pure gave us four practical benefits. Policies are unit-testable, and the policy suite requires coverage of every production module. A ``what-if'' admission query can preview a proposed change without touching live state. The policy and organization data are distributed as a versioned bundle (section~\ref{sec:control}), so every data-plane replica evaluates the same revision. And the migration itself was safe: the previous Go engine and the OPA engine ran side by side behind a selector while a parity suite checked identical decisions across all combinations of engine and enforcement mode, together with hot-reload tests of updated rules.

\begin{figure*}[t]
\centering
\resizebox{\textwidth}{!}{%
\begin{tikzpicture}[font=\small]
\tikzset{lc/.style={minimum width=3.4cm, minimum height=1.45cm}}
\node[obox, lc] (a1) at (0,0)    {\textbf{Author}\\[-1pt]{\footnotesize platform or BU admin}\\[-2pt]{\footnotesize (SSO-derived role)}};
\node[obox, lc] (a2) at (3.8,0) {\textbf{Validate}\\[-1pt]{\footnotesize schema, conflicts,}\\[-2pt]{\footnotesize locks, dry-run}};
\node[obox, lc] (a3) at (7.6,0)  {\textbf{Approve}\\[-1pt]{\footnotesize BU changes queued}\\[-2pt]{\footnotesize or auto-approved;}\\[-2pt]{\footnotesize no self-approval}};
\node[obox, lc] (a4) at (11.4,0){\textbf{Schedule}\\[-1pt]{\footnotesize activate at local}\\[-2pt]{\footnotesize midnight, or now}};
\node[obox, lc] (a5) at (15.2,0) {\textbf{Persist + audit}\\[-1pt]{\footnotesize durable store, epoch,}\\[-2pt]{\footnotesize activity log}};
\node[bbox, lc] (a6) at (19.0,0){\textbf{Project}\\[-1pt]{\footnotesize leader-elected,}\\[-2pt]{\footnotesize deterministic bundle}};
\node[bbox, lc] (b1) at (19.0,-2.0){\textbf{Distribute}\\[-1pt]{\footnotesize versioned; conditional}\\[-2pt]{\footnotesize polling; hot reload}};
\node[ebox, lc] (b2) at (15.2,-2.0) {\textbf{Enforce}\\[-1pt]{\footnotesize every request;}\\[-2pt]{\footnotesize last-good on failure}};
\node[ebox, minimum width=6.6cm, minimum height=1.15cm] (b3) at (9.1,-2.0) {\textbf{Rollout modes}\\[-1pt]{\footnotesize tracking always on $\rightarrow$ \emph{shadow} (evaluate,}\\[-2pt]{\footnotesize log, never block) $\rightarrow$ \emph{authoritative} (enforce)}};
\node[gbox, minimum width=6.4cm, minimum height=1.15cm] (b4) at (1.5,-2.0) {\textbf{Feedback}\\[-1pt]{\footnotesize utilization \%, threshold emails, status lines,}\\[-2pt]{\footnotesize leadership digests, saved vs.\ enforced revision}};
\draw[arr] (a1)--(a2); \draw[arr] (a2)--(a3); \draw[arr] (a3)--(a4); \draw[arr] (a4)--(a5); \draw[arr] (a5)--(a6);
\draw[arr] (a6)--(b1); \draw[arr] (b1)--(b2); \draw[arr] (b2)--(b3); \draw[arr] (b3)--(b4);
\draw[darr] (b4.north -| a1) -- node[right, font=\scriptsize\itshape, text=black!60]{inform next change} (a1.south);
\end{tikzpicture}}
\caption{Policy lifecycle. Intent is authored and validated against the same rules the data plane enforces, approved when it originates from a business unit, persisted with an audit record, projected into a versioned bundle, and enforced on every request. New controls are rolled out in shadow mode before they block.}
\label{fig:lifecycle-policy}
\end{figure*}
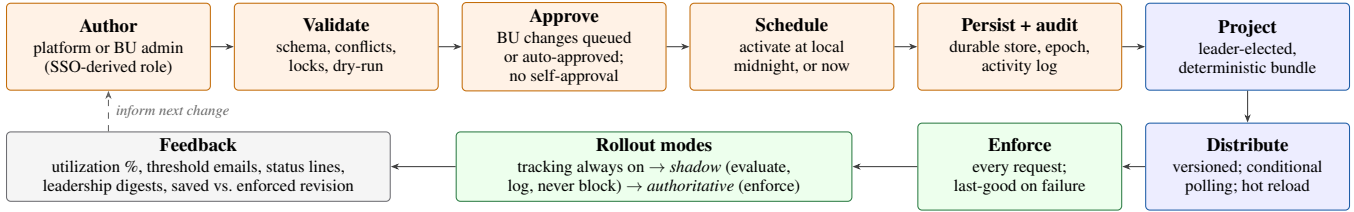

\section{The Governance Control Plane}\label{sec:control}
A policy engine is useful only if the right people can change policy safely. The control plane turns models, prices and policy from engineering artifacts into an operated product (fig.~\ref{fig:lifecycle-policy}).

\para{Roles and trust.} Administrators authenticate through corporate single sign-on, and the gateway maps each administrator onto one of two canonical roles. \emph{Platform administrators} act organization-wide. \emph{BU administrators} act only within the business units they have been granted.

\subsection{Model onboarding lifecycle}\label{sec:onboarding}
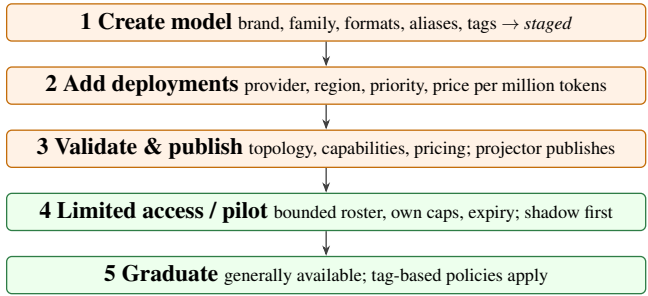
\begin{figure}[t]
\centering
\resizebox{\columnwidth}{!}{%
\begin{tikzpicture}[font=\small]
\tikzset{st/.style={minimum width=8.6cm, minimum height=0.5cm}}
\node[obox, st] (s1) at (0,0)     {\textbf{1 Create model} {\scriptsize brand, family, formats, aliases, tags $\rightarrow$ \emph{staged}}};
\node[obox, st] (s2) at (0,-0.85) {\textbf{2 Add deployments} {\scriptsize provider, region, priority, price per million tokens}};
\node[obox, st] (s3) at (0,-1.7)  {\textbf{3 Validate \& publish} {\scriptsize topology, capabilities, pricing; projector publishes}};
\node[ebox, st] (s4) at (0,-2.55) {\textbf{4 Limited access / pilot} {\scriptsize bounded roster, own caps, expiry; shadow first}};
\node[ebox, st] (s5) at (0,-3.4)  {\textbf{5 Graduate} {\scriptsize generally available; tag-based policies apply}};
\draw[arr] (s1)--(s2); \draw[arr] (s2)--(s3); \draw[arr] (s3)--(s4); \draw[arr] (s4)--(s5);
\end{tikzpicture}}
\caption{Model onboarding through the control plane: each step is a validated, audited data change; no code release is involved.}
\label{fig:onboard}
\end{figure}

Onboarding a model is a sequence of validated data changes through the management API (fig.~\ref{fig:onboard}). A platform administrator creates the model (brand, API family, entry formats, aliases, tags, capabilities) and then its deployments, each with provider, region, priority and price. Every write is validated against the registry topology, capability and pricing rules; updates and deletes require an entity-tag precondition (missing returns 428, stale returns 409), and an idempotency key makes retries safe. Until it has an enabled deployment, a model is \emph{staged} and not served. A new model can first be released as \emph{limited access} to a bounded named roster, with its own spend windows, amounts and expiry, or through a pilot. It is then \emph{graduated}, which clears the roster and makes the model generally available. Because policies can select models by tag, a newly tagged model is covered by existing restrictions and budgets within one publication cycle, with no policy edit. A model cannot be deleted while a live policy references it, and every step is recorded in the activity log.

\subsection{Platform and business-unit policies}\label{sec:bupolicies}
\para{Platform policies} set organization-wide controls: per-person runaway-cost limits, per-model caps, model-access decisions, and \emph{locks} that BU policies cannot exceed. Conflict checks scan the complete active set before a change is accepted, and caps must be strictly ordered across windows (a monthly cap exceeds the weekly cap, which exceeds the daily cap).

\para{BU policies} let business-unit administrators set per-person and per-model caps and model toggles for their own organization, within the platform's locks: an unlocked BU cap takes precedence over the platform baseline, a request above a locked maximum is rejected (and any stored cap is clamped to that maximum when policies are applied), and an exact lock rejects any change. BU model toggles take effect without approval but cannot re-enable a model the platform has locked. Controls can select models explicitly or by tag, and tag selections follow the registry.

\para{Approvals and scheduling.} BU cap changes are either submitted to an approval queue, where they do not affect enforcement until approved, or auto-approved where the platform allows it. Decisions are restricted to platform administrators, self-approval is forbidden, a rejection must include a comment, and an approval can be bound to the reviewed diff so that edits arriving in between are detected. Changes can be scheduled to activate at midnight in a given timezone. A leader-gated scheduler applies them, expires stale requests, and re-projects whenever the store's epoch is ahead of the served bundle, so the durable record and the enforced bundle converge.

\para{Pilots.} New models and new budget regimes are introduced through \emph{pilots}: named, date-bounded programs of a small number of cohorts. Each cohort has its own cost controls, model-access lists, BU assignment, alerts and exceptions. Cohorts that share a leader and window but have distinct amounts form a \emph{tier ladder}. An enrolled user is governed by the lowest tier whose exceptions do not cover them, so an exception promotes a user up the ladder rather than removing the cap. Validation returns every violation at once (double enrollment, duplicate controls, duplicate tier amounts, unknown or empty tag selections), and writes serialize across replicas under a database advisory lock. Pilot enforcement has its own switch: \emph{off}, \emph{shadow} (evaluate and log only) or \emph{authoritative}.

\subsection{Usage notifications and reminders}\label{sec:notify}
Users should learn that they are approaching a limit long before they hit it. Notification points are configurable at two levels. In the rule language, any entry can attach an email or status-line action at its own threshold; the typical pattern is a warning email at 90\% of an allowance and a block plus email at 100\% (appendix~\ref{app:rules}). In managed policies, each control carries alerts at administrator-chosen percentages (1--100\%), each with at least one channel (email or in-tool message) and a template. Templates are managed through the API, are scoped to the platform or to a business unit, and support placeholders such as utilization percentage, threshold, resume time, model, window, person and business unit. Delivery is deduplicated per policy owner, window instance, threshold, template and channel, so a user receives one email per crossing rather than one per request. Status headers on every response report the governing limit's utilization, status, scope, reset time and any pause, and coding agents and the developer CLI render them as a live status line. Leadership notifications are deferred into one digest per leader after the window boundary. The usage API reports utilization as percentages.

\subsection{Rollout, distribution and audit}\label{sec:rollout}
\para{Shadow first.} Usage tracking is always on, and enforcement is a separate switch, set globally and per pilot. New controls are therefore rolled out in shadow mode against real traffic, with utilization visible to users and administrators and nothing blocked, until their effect is understood. Because scopes are views over current-window counters, a draft rule can also be dry-run against current usage before it is enabled. A rule enabled directly while enforcement is on applies to the window to date and can block at once, which is why new controls go through shadow mode or a pilot first.

\para{Distribution.} The durable store is the system of record. A leader-elected projector builds a deterministic bundle containing policy, registry-derived access data and the organization snapshot, and serves it with an entity tag. Data-plane replicas poll conditionally and hot-reload. A failed reload keeps the last good revision, and an engine whose initial bundle fails to compile refuses to serve. API responses distinguish the \emph{saved} revision from the \emph{enforced} revision, so administrators can see when a change is live.

\para{Audit.} Every mutation writes an append-only activity record; policy changes write it in the same transaction as the change. Records can be filtered by category, outcome, actor and role; platform administrators see all records, and BU administrators see their own business units.

\section{Telemetry, Metering and Attribution}\label{sec:telemetry}
\para{Metrics and traces.} Envoy's native statistics (upstream health, circuit breakers, connections, latency and response classes) and application metrics (requests, latency, token counts by kind, cache reads and writes, cost, rate-limit events, and schema-drift counts) are exported through OpenTelemetry~\cite{otel}. Labels identify the canonical model, vendor and entry path, and aliases collapse to canonical identifiers so that each logical model is counted once. Per-caller dimensions are limited to a small set of governance events (such as limit blocks); full per-request attribution lives in the usage record, which keeps metric cardinality bounded and identity data in an access-controlled store.

\para{The access log as the system of record.} We initially emitted usage events from the control service through an asynchronous in-memory queue. That design lost data silently: when the queue was full, after retries were exhausted, during pod shutdown, and when a client disconnected before the trailers arrived. We moved the durable record to Envoy's access log, which Envoy writes for \emph{every} stream, including resets, immediate rejections and control-service outages. The control service contributes attribution (user, application, tier, platform, workflow, harness, requested and canonical model, entry and upstream formats) and the priced cost breakdown as metadata assembled from in-memory state only. Assembling this metadata adds no I/O to any ext\_proc phase. The Wasm filter contributes per-field token counts as trailers. Interrupted streams still produce a usage record. Blocked requests (4xx immediate responses for authentication failures, invalid requests, access denials and limits) carry the same attribution, which makes rejected traffic visible to abuse and chargeback analysis. The earlier event pipeline was run in parallel during migration so the two could be reconciled.

\para{From log line to warehouse.} Each record is one JSON line on standard output. A node-level logging agent ships it to a managed log service, from which the records are routed to the analytics warehouse. Rows are projected into a typed usage table, partitioned by day and clustered by caller and model, and a view deduplicates on the gateway-generated request ID, which clients cannot spoof. From the logging agent onward, delivery is at-least-once with read-time deduplication, barring node loss. Once a line reaches standard output it sits on node disk under the container runtime's management, so a crash of the gateway process afterwards does not lose it. The residual loss window is bounded: a hard proxy crash can lose at most one flush interval of buffered entries (below), and a node failure can lose seconds of unshipped agent buffer. This replaced the unbounded, silent loss modes of the earlier in-process queue. The schema is additive-only and versioned, so new fields require no infrastructure change. These records, one per request with tokens and cost by component, drive chargeback and usage analytics. Appendix~\ref{app:usage} lists the fields.

\para{Implementation details that mattered.} First, Envoy silently \emph{drops} dynamic metadata returned by an ext\_proc service unless the filter allow-lists the receiving namespaces. Without the allow list, every metadata-derived column is null, and only an end-to-end test against a real Envoy caught it. Second, Envoy is configured to flush access-log entries every few seconds for streams that are still open, rather than only when a stream closes, which bounds what a hard proxy crash can lose on a long stream at negligible I/O cost. Periodic entries preserve the record and its attribution; usage is carried by the entry written when the stream ends. Third, Envoy is configured not to preserve client-supplied request IDs, so a client cannot poison the deduplication key.

\para{Context-engineering signals.} The Wasm filter reports which tools a model actually invoked. Combined with the tools a client declared, this lets the platform identify MCP servers and tools that a user keeps declaring across requests within a 24-hour session window but that the model never invokes, and return per-user recommendations with the estimated token savings per request of removing them.

\section{Reliability and Real-Time Traffic}\label{sec:reliability}
\para{Voice changed the problem.} Our previous gateway was built for short request--response exchanges behind a request-oriented L7 load balancer. Chat traffic suited it, because each request released its resources almost immediately. A voice call is different: one WebSocket held for the whole conversation and streaming audio in both directions. Two thousand concurrent calls of up to 30 minutes each pinned per-session resources for the life of the call in a service sized for sub-second requests. In addition, a load balancer that balances \emph{requests} is a poor fit for long-lived \emph{connections}. Envoy resolves both problems: a call becomes a connection multiplexed on Envoy's per-worker event loops, and connection lifecycle is a first-class proxy concept.

\para{A connection-oriented path.} Real-time routes are tuned for long-lived sessions: frames are forwarded as they arrive, with no body buffering and no request timeout, and an idle timeout reclaims abandoned connections. Real-time sessions are recorded in the same access log and proxy metrics as other traffic, and real-time model endpoints are available in more than one region.

\para{Graceful, drain-aware deployments.} Pods move for ordinary reasons: deploys, autoscaling and node upgrades. A chat request that dies in a rollout is a retry; a voice call that dies is a customer hung up on mid-sentence. Our shutdown sequence is designed around the longest call:
\begin{enumerate}[leftmargin=1.5em]
\item A pre-stop hook marks the proxy unhealthy through its admin interface, so the load balancer stops sending it new connections while existing ones continue.
\item The hook then polls the listener's active downstream connection count periodically and returns as soon as it reaches zero, or shortly before the grace period ends.
\item The termination grace period (one hour) is longer than the longest expected call, because a conversation can last 30 minutes and holding a pod open is preferable to cutting off a customer.
\item Rollouts add a new proxy pod before removing an old one (surge one, zero unavailable), a pod disruption budget keeps all but one of the minimum replicas available during voluntary disruptions, and the proxy tier autoscales horizontally in each region.
\item The gRPC services drain the same way: they report \texttt{NOT\_SERVING} on their health endpoint and then stop gracefully within a shorter grace period of their own. Their rollouts surge a full replica set with zero unavailability, because a fail-closed dependency must finish draining before it stops.
\end{enumerate}
With this sequence we have deployed more than ten times during live voice traffic, with zero dropped calls, while sustaining more than 2,000 concurrent calls.

\para{Regional placement and failover.} The gateway runs in multiple regions with disaster recovery and load balancing. Where a data-residency endpoint is configured for a vendor, it is the default. Automatic fallback to another endpoint is a single retry, limited to connection-level failures and service-unavailable responses; other errors are returned to the client rather than retried, which keeps retries from amplifying load during a vendor incident.

\para{Migration without a big bang.} The gateway reached its current form without asking any team to change its integration. First, Envoy was placed in front of the existing service, so nothing behind it changed. Then authentication, credential brokering, translation, spend and policy moved into the new data plane one capability at a time, behind the same entry point, following a strangler-fig pattern~\cite{strangler}. Progress was measured continuously, and the previous implementation is being retired capability by capability.

\section{Evaluation}\label{sec:eval}
\subsection{Gateway overhead}
\para{Method.} We measured the latency the gateway adds with a Locust load generator (user ramp of 500 users/s, think time 0.5--1\,s) against a synthetic upstream with a fixed 110\,ms latency. Each instance has 4 vCPU and 8 GiB, split evenly between one Envoy pod and one control-service pod. Load steps use 1,000 and 2,000 virtual users. Every request carries a unique payload, and each load step runs for five minutes. \emph{Overhead} is the time from the start of the client request to the start of the upstream response, minus the upstream service time, both recorded by Envoy at microsecond resolution; it therefore measures the request path up to the upstream call and excludes response-body work. We ran two and four instances with 1,000 and 2,000 virtual users (about 1,100 and 2,300 requests per second) against the unified chat-completions path.

\begin{table}[t]
\centering\footnotesize
\begin{tabular}{@{}rrrrrr@{}}
\toprule
 & & \multicolumn{3}{c}{\textbf{Overhead (ms)}} & \\
\cmidrule(lr){3-5}
\textbf{Instances} & \textbf{RPS} & \textbf{p50} & \textbf{p95} & \textbf{p99} & \textbf{Failed}\\
\midrule
2 & 1,137 & 4 & 10 & 11 & 0\\
2 & 2,257 & 4 & 10 & 11 & 0\\
4 & 1,138 & 4 & 10 & 11 & 0\\
4 & 2,262 & 4 & 9 & 10 & 0\\
\bottomrule
\end{tabular}
\caption{Gateway overhead (time to the start of the upstream response, minus upstream service time) with a synthetic 110\,ms upstream. RPS: achieved requests per second.}
\label{tab:overhead}
\end{table}

\begin{figure}[t]
\centering
\begin{tikzpicture}
\begin{axis}[
  width=\columnwidth, height=3.8cm,
  ybar,
  ymin=0, ymax=14, ytick={0,5,10}, bar width=6pt,
  ylabel={overhead (ms)}, ylabel style={font=\scriptsize},
  symbolic x coords={p50,p95,p99}, xtick=data,
  enlarge x limits=0.25,
  tick label style={font=\scriptsize},
  nodes near coords, nodes near coords style={font=\tiny},
  legend style={font=\scriptsize, at={(0.5,1.04)}, anchor=south, legend columns=2, draw=none, fill=none, /tikz/every even column/.append style={column sep=4pt}},
  legend image code/.code={\draw[#1] (0cm,-0.08cm) rectangle (0.12cm,0.12cm);},
  axis x line*=bottom, axis y line*=left,
]
\addplot[fill=cblue!90!black, draw=cblue!90!black] coordinates {(p50,4) (p95,10) (p99,11)};
\addplot[fill=cblue!40, draw=cblue!40] coordinates {(p50,4) (p95,10) (p99,11)};
\addplot[fill=cgreen!80, draw=cgreen!80] coordinates {(p50,4) (p95,10) (p99,11)};
\addplot[fill=green!25, draw=green!35] coordinates {(p50,4) (p95,9) (p99,10)};
\legend{2 inst @1.1k RPS, 2 inst @2.3k RPS, 4 inst @1.1k RPS, 4 inst @2.3k RPS}
\end{axis}
\end{tikzpicture}
\caption{Overhead percentiles by instance count and load. Overhead was essentially unchanged when load or instance count doubled within the tested range.}
\label{fig:overhead}
\end{figure}
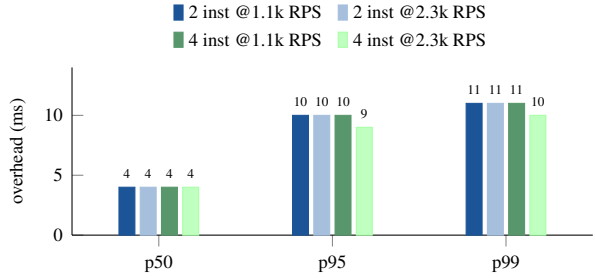

\para{Results.} Table~\ref{tab:overhead} and fig.~\ref{fig:overhead} show two results. First, overhead remained \emph{stable over the tested range}: doubling the request rate on two instances changed no percentile, and on four instances the 2,000-user figures were within 1\,ms of the 1,000-user figures. Second, adding instances did not change overhead, which is consistent with a per-request cost rather than queueing delay at these rates. We did not test beyond 2,262 RPS and make no claim about behavior near saturation. No step produced a failed request.

\para{Threats to validity.} The benchmark measures the core request-handling path (caller authentication, model resolution, credential brokering and request rewriting) against a synthetic upstream; production overhead varies with the policies and features enabled for a workload. The load generator ran inside the same cluster, and all virtual users shared one credential, so caller authentication was served from the control service's in-process cache. The metric excludes response-body processing, which includes streaming translation. We make the methodology explicit so that others can reproduce it.

\subsection{Production operation}
Table~\ref{tab:prod} summarizes production operation as of September 2026. All figures are historical, point-in-time aggregates, not service-level commitments.

\begin{table}[t]
\centering\footnotesize
\setlength{\tabcolsep}{2pt}
\begin{tabular}{@{}lr@{}}
\toprule
\textbf{Measure} & \textbf{Value}\\
\midrule
Developers using the gateway & 12,000+\\
Production use cases & hundreds\\
Requests per day & millions\\
Tokens per day (all traffic) & $\sim$200 billion\\
Models / model families / providers & 100+ / multiple / multiple\\
Peak throughput from a single workload & 60,000 req/min\\
Concurrent real-time voice calls supported & 2,000+\\
Deploys during live voice traffic; dropped calls & 10+; 0\\
Gateway prompt-cache hit rate (best workloads) & up to 96\%\\
Availability to business (ATB) & 99.999\%\\
\bottomrule
\end{tabular}
\caption{Production operation (historical point-in-time figures as of September 2026, not service-level commitments). ATB is PayPal's availability-to-business service-level indicator.}
\label{tab:prod}
\end{table}

\para{Effect of centralization.} Several outcomes were not achievable under direct integration. Governed AI traffic now flows through one mediated entry point, and model calls carry per-user or per-application, per-platform and per-workflow attribution and a priced usage record. Model choice is configuration: switching the vendor or region behind a model requires no application change, and onboarding a new model or changing a price is a validated data change. Budgets are organization-aware, shadow-testable and retroactive within their window, and users are typically warned before they reach a limit. Engineers reach approved models through their supported entry formats with no vendor keys to manage and no per-team provisioning. Gateway-level prompt-cache shaping reached hit rates of up to 96\% on agentic coding workloads, which lowers input cost, since cache reads are billed at a discount~\cite{anthropiccache}, and can reduce time to first token, with no client changes.

\subsection{Engineering assurance}
The gateway is security- and cost-critical, so we treat its correctness as a product requirement. Continuous integration runs the Go and Rust test suites with a coverage floor, the policy test suite (which requires coverage of every production policy module), bundle validation, and the cross-language contract tests for the handoff, together with a build-failing coverage test for codec field classification. An extended suite adds database-backed tests for the control-plane stores and engine-parity tests between the previous Go engine and the OPA engine. A synthetic upstream service lets end-to-end tests and benchmarks run without vendor spend. The data plane, control plane and identity services comprise roughly 45k lines of Go and 7k lines of Rust, with about 35k further lines of Go tests.

\section{Lessons Learned}\label{sec:lessons}
\begin{enumerate}[label=\textbf{L\arabic*.}, leftmargin=2.2em, labelsep=0.4em, align=right]
\item \textbf{Wrap, don't rewrite.} Putting Envoy in front of a working service and moving capabilities behind it one at a time required no client migrations, with no regressions observed during the monitored rollout. Measuring the share of traffic on the new path made progress visible.
\item \textbf{Split control from data by latency profile.} Out-of-process for I/O-bound decisions and in-process for per-byte work let us be strict and fast at the same time. The same proxy foundation is what made voice scale: long-lived connections are multiplexed on Envoy's per-worker event loops rather than pinning per-session resources.
\item \textbf{Treat the connection lifecycle as a first-class design input.} An HTTP request can be handed off and forgotten; a WebSocket must be held and then drained. Upgrade, stream, idle and drain shaped most of our reliability decisions.
\item \textbf{Make models, prices and policy data, and validate them like code.} Data-driven configuration is safe only with strict validation. A gateway that refuses to start on bad configuration is safer than one that starts and routes incorrectly.
\item \textbf{Store usage at the finest grain, keyed by immutable identifiers, and make every budget a view.} This decision removed the need for data migrations when budgets or tags change, enabled rules that apply to the window to date, and made ``what would this rule do now?'' answerable.
\item \textbf{Keep decisions pure.} Separating gather, decide and effect made policy testable and previewable, and made it possible to migrate engines under a parity harness.
\item \textbf{Ship governance in shadow mode, and warn before you block.} Always-on tracking with separate enforcement switches let new controls prove themselves on real traffic before they blocked anyone, and threshold reminders gave users advance notice before limits applied.
\item \textbf{Anchor policy to stable identities.} Rules that name organization leaders rather than team lists survive reorganizations without edits.
\item \textbf{Put the durable record where it is hardest to lose.} Moving attribution from an in-application queue to the proxy's access log replaced silent, unbounded loss with a small, bounded loss window and made usage a system of record.
\item \textbf{Routing is not the differentiator.} Off-the-shelf proxies route well. What our organization needed, and the off-the-shelf proxy we evaluated did not provide at the time, was real-time support, identity-aware authorization, organization-aware spend control and workload-identity-based cloud credentials.
\end{enumerate}

\section{Limitations and Future Work}
Several limitations remain. Centralizing access also concentrates risk: the gateway brokers credentials for every provider, so it is a high-value target whose own access, secret handling and change control must meet the standard of the most sensitive workload it serves. Our experience comes from one large organization, and other environments may weigh these trade-offs differently. Our overhead benchmark covers the core request path, and production overhead depends on the policies and features enabled for a workload. Data-loss prevention relies on a learned detector whose accuracy varies by entity type, so it reduces rather than eliminates exposure. Budget thresholds, notification points and data-protection policies are organization-specific and require ongoing tuning against real usage; the mechanisms described here make that tuning safe, but do not remove the need for it.

Work in progress includes: a single credential for reaching every model, with automatic model fallbacks; additional providers; extending the same entry point to MCP tool calls and autonomous agents, with identity, policy and audit applied to every action; further safety and quality controls at the gateway; cost- and latency-aware routing and semantic caching, building on work such as~\cite{frugalgpt,routellm}; and richer, more proactive spend insights for owners.

\section{Conclusion}
The Cosmos LLM Gateway shows that, in one large enterprise, the problems of LLM access (credential sprawl, format lock-in, ungoverned spend and uneven governance) could be addressed at a single layer while adding a median of a few milliseconds of measured request-path overhead (section~\ref{sec:eval}) for the developers and products that depend on it. The architecture rests on a few choices: governance out of process and translation in process; models, prices and policy as validated data; spend scopes as views over counters keyed by immutable model identifiers; pure policy decisions distributed as versioned bundles and introduced in shadow mode, with users normally warned before they are blocked; and a durable, priced usage record written where it is hardest to lose. Viewed this way, an enterprise LLM gateway is neither just a router nor just a policy engine. It is the platform layer that absorbs vendor multiplicity, credential variability and format heterogeneity into one governed substrate, so that a developer writes against one API, presents one identity, and reaches the right model through its supported entry formats with governance, attribution and observability applied by default.

\section*{Acknowledgments}
We thank the PayPal Cosmos AI platform engineering, product and program teams, and the many internal teams who adopted the gateway early and shaped it with their feedback. All operational figures are historical, point-in-time aggregates; they are not service-level commitments and do not constitute forward-looking statements. The architecture described here provides technical controls only and does not by itself establish legal or regulatory compliance. Product and project names are trademarks of their respective owners.

\onecolumn
\appendix

\vspace*{-1.5ex}
\section{Registry Excerpt}\label{app:registry}
Abridged and sanitized. The model identifier is illustrative and prices are placeholders (USD per million tokens). Provider, credential and path configuration are omitted.
\begin{lstlisting}[style=code,language=yamlish,basicstyle=\ttfamily\fontsize{6.6pt}{7.3pt}\selectfont]
families:     { chat: {}, responses: { canonical_format: openai_responses }, embeddings: {} }
wire_formats: { anthropic_messages: { family: chat }, openai_chat: { family: chat },
                anthropic_vertex: { family: chat }, gemini_generate_content: { family: chat },
                openai_responses: { family: responses } }
converters:
  - { from: anthropic_messages, to: openai_chat }
  - { from: anthropic_messages, to: anthropic_vertex }
  - { from: anthropic_messages, to: openai_responses }
  - { from: anthropic_messages, to: gemini_generate_content }
  - { from: openai_chat,        to: anthropic_vertex }
  - { from: openai_chat,        to: openai_responses }
  - { from: openai_chat,        to: gemini_generate_content }
models:
  claude-opus-x:                                   # illustrative model id
    brand: anthropic
    entry_formats: [anthropic_messages, openai_chat]
    aliases: [claude-opus-x-latest]
    tags: [all, claude, opus, proprietary, frontier]
    deployments:
      - id: opus-x@cloud
        provider: <cloud-ai-platform>
        region: global
        deployment_format: anthropic_vertex
        priority: 1
        pricing: { unit: usd_per_mtok, input: <in>, output: <out>,
                   cache_read: <cr>, cache_write: <cw>, cache_write_1h: <cw1h>,
                   threshold_tokens: <n>,                       # optional long-context tier
                   long: { input: <in2>, output: <out2>, cache_read: <cr2>, cache_write: <cw2> } }
\end{lstlisting}

\section{Spend Rule Language}\label{app:rules}
Each rule holds one or more entries; each entry reads as one sentence: \emph{when} a user's spend in a scope crosses a threshold within a window, \emph{do} these actions. Identifiers, weights, amounts and template names are illustrative placeholders, not production values. \texttt{scope} is a registry tag or a single model identifier and is resolved at evaluation time (section~\ref{sec:spend}). Rules default to \emph{enabled}, because a cap someone forgot to enable silently does nothing, which is the worse failure for cost control. Whether an enabled rule can block is governed separately by the enforcement mode (section~\ref{sec:rollout}). The first rule below shows the reminder pattern: a warning email at 90\% of an allowance, then a block and a second email at 100\%.
\begin{lstlisting}[style=code,language=yamlish,basicstyle=\ttfamily\fontsize{6.6pt}{7.3pt}\selectfont]
rules:
  - id: frontier-family-monthly          # stable id; surfaces in metrics
    weight: 60                           # precedence (higher first)
    matches:
      - match: { tiers: [normal], scope: opus, window: monthly, at_or_above_usd: <0.9 x T> }
        actions:
          - { type: email, to: [user], subject: "You're at 90% of your monthly limit", body_file: <template> }
      - match: { tiers: [normal], scope: opus, window: monthly, at_or_above_usd: <T> }
        actions:
          - { type: email, to: [user], subject: "Monthly limit reached", body_file: <template> }
          - { type: block }              # stateless; clears when the window rolls
  - id: daily-pause
    weight: 50
    matches:
      - match: { scope: all, window: daily, at_or_above_usd: <D> }
        actions:
          - { type: pause, hours: <H> }  # one-shot per (rule, window); self-lifts
          - { type: message, level: warn, text: "Paused. Resumes {resume_time}." }
          - { type: email, to: [user] }
  - id: org-monthly-cap
    weight: 100
    matches:
      - match: { tiers: [normal, pro], scope: all, window: monthly, leaders: [<org-leader>],
                 exempt_users: [<login>], at_or_above_usd: <B> }
        actions:
          - { type: block }
          - { type: message, level: warn, text: "You have reached your organization's monthly per-person limit." }
  - id: weekly-leadership-awareness
    weight: 40
    matches:
      - match: { tiers: [normal, pro], scope: all, window: weekly, at_or_above_usd: <W> }
        actions:
          - { type: email, to: [leadership] }   # deferred digest after the window boundary
\end{lstlisting}

\section{Usage Record}\label{app:usage}
One row per request in a day-partitioned warehouse table, clustered by caller and model, read through a view that deduplicates on the request ID.

\smallskip
{\footnotesize
\noindent\begin{tabularx}{\textwidth}{@{}lXl@{}}
\toprule
\textbf{Field group} & \textbf{Fields} & \textbf{Origin}\\
\midrule
Identity of the request & request id (gateway-generated), timestamp, trace id & Envoy\\
Caller & user, application, tier, platform/workflow/harness tags (one JSON object) & control (request phase)\\
Routing & requested model, canonical model, vendor, entry and upstream formats, resource type & control (request phase)\\
Usage & input, output, cache-read, cache-write (incl.\ extended-lifetime) tokens & Wasm trailers\\
Cost & per-component and total USD; workflow spent/limit when present & control (trailer phase)\\
Outcome & status, response flags, duration, time to first byte, upstream cluster & Envoy\\
\bottomrule
\end{tabularx}
}

\end{document}